\documentclass[a4paper,11pt]{article}
\pdfoutput=1 

\usepackage{jcappub} 

\usepackage[T1]{fontenc} 

\usepackage{amsfonts}
\usepackage{amsmath}
\usepackage{amssymb}
\usepackage{graphicx,color}
\usepackage{float}
\usepackage{hyperref}
\usepackage{subfigure}
\usepackage{dcolumn}
\usepackage{soul}
\usepackage{ulem}
\usepackage{verbatim}

\title{Thermal effects on warm chromoinflation}

\author[a,b,c,1]{Vahid Kamali}
\author[d,2]{and Rudnei O. Ramos \note{Corresponding author.}}

\affiliation[a]{Department of Physics, McGill University, Montr\'{e}al,
	QC, H3A 2T8, Canada}
\affiliation[b]{Department of Physics, Bu-Ali Sina (Avicenna) University, Hamedan 65178,
	016016, Iran}
\affiliation[c]{School of Physics, Insitute for Research in Fundamental Sciences (IPM),
	19538-33511, Tehran, Iran}
\affiliation[d]{Departamento de F\'{\i}sica Te\'orica, Universidade do Estado do Rio de Janeiro, 
20550-013 Rio de Janeiro, RJ, Brazil}

\emailAdd{vkamali@ipm.ir}
\emailAdd{rudnei@uerj.br}

\abstract{

We explore a model of a pseudo-Nambu-Goldstone boson inflaton field
coupled to a non-Abelian $SU(2)$ gauge field. This model naturally
leads to a warm inflation scenario, where the inflationary dynamics is
dominated by thermal dissipation.  In this work, we consider a
scenario where the inflaton, an axion-like field, is coupled to the
$SU(2)$ gauge field, similar to chromoinflation models. Both the
inflaton and the gauge field with a non-vanishing vacuum expectation
value are coupled to a thermal radiation bath.  We demonstrate that
the presence of the thermal bath during warm chromoinflation induces a
thermal plasma mass for the background gauge field. This thermal mass
can significantly disrupt the dynamics of the background gauge field,
thereby driving it to its trivial null solution.
 
}

\begin{document}
\maketitle

\section{Introduction}

Warm inflation~\cite{Berera:1995wh,Berera:1995ie,Berera:1998px}
has recently seen successful implementations in models featuring
pseudo-Nambu-Goldstone scalar
fields~\cite{Bastero-Gil:2016qru,Bastero-Gil:2019gao,Berghaus:2019whh,Laine:2021ego,DeRocco:2021rzv}.
These models often involve axion-like fields directly coupled to non-Abelian
gauge fields. The dissipation of the axion-inflaton into gauge fields
can naturally lead to a thermal radiation bath, even from initial
vacuum conditions~\cite{Cheng:2015oqa,Laine:2021ego,DeRocco:2021rzv}. This thermal
bath is a hallmark of the warm inflationary regime.

The explicit realization of consistent warm inflation dynamics in
these models is a significant achievement. Warm inflation has emerged as a
promising inflation model that aligns with effective field theory and
potentially finds a ultraviolet (UV) completion in quantum
gravity~\cite{Das:2018rpg,Motaharfar:2018zyb,Goswami:2019ehb,Berera:2019zdd,Kamali:2019xnt,Berera:2020iyn,Das:2020xmh,Brandenberger:2020oav,Motaharfar:2021egj}.
Therefore, constructing explicit models of warm inflation and
exploring their implications has become increasingly important.

In this work, we investigate an axion warm inflation model
incorporating a Chern-Simons interaction between the axion inflaton
field and a $SU(2)$ gauge field. It is
well-established~\cite{Maleknejad:2011jw,Maleknejad:2016qjz,Adshead:2012kp,Kamali:2019ppi}
that, in addition to the background inflaton field, an isotropic
solution for the $SU(2)$ gauge field is also permissible (for a review,
see, e.g. ref.~\cite{Maleknejad:2012fw}). This background gauge field
can substantially influence the evolutionary dynamics of the expanding
system at both the background and perturbation levels.

Here, we demonstrate that the presence of a thermal bath naturally
induces a plasma mass for the background gauge field. This thermal
mass contribution to the gauge background field, similar to the
thermal mass affecting the inflaton equation in warm
inflation~\cite{Berera:1998gx,Yokoyama:1998ju}, can potentially
disrupt the slow-roll conditions for the gauge field, driving it
towards a vanishing value. We investigate the impact of these thermal
effects on the background dynamics and illustrate our results using a
commonly employed axion potential.

This paper is organized as follows. In section~\ref{sec2}, we briefly
review the background dynamics of chromoinflation in the context of
warm inflation. In section~\ref{sec3}, we compute the leading order
thermal contributions to the background dynamics.  In section~\ref{sec4},
we illustrate our results by explicitly studying the dynamics in a
case of axion-like potential.  In section~\ref{sec5}, we give our
concluding remarks.

Throughout this paper, we work with the natural units, in which the
speed of light,  Planck's constant and Boltzmann's constant are all
set to $1$, $c=\hbar=k_B=1$.  We
work in the context of a spatially flat homogeneous and isotropic
background metric with scale factor $a(t)$, where $t$ is physical
time. We also work with the reduced Planck mass, defined as
$M_{\rm Pl} = (8\pi G)^{-\frac{1}{2}} \simeq 2.44\times 10^{18}$ GeV 
and where $G$ is Newton's gravitational constant.
The Hubble expansion rate is
$H(t) = {\dot{a}(t)}/a(t)$, where an overdot denotes the derivative
with respect to time.

\section{The Axion Warm Inflation Model}
\label{sec2}

We work with a minimal setting for an axion-like field $\phi$ making
the role of the inflaton and coupled to a non-Abelian $SU(2)$ gauge
field $A_\mu$ with the standard dimension five interaction, with
action in a FLRW metric,
\begin{eqnarray}
S &=& \int d^4 x a^3(t) \left[\frac{1}{2}(\partial_\mu \phi)^2 -
  V(\phi) -\frac{1}{4}F_{\mu\nu}^cF^{c \,\mu \nu} \right.  \nonumber
  \\ &-& \left. \frac{\lambda}{4f}\phi  F_{\mu \nu}^c \tilde{F}^{c\,
    \mu \nu}\right],
\label{lagr}
\end{eqnarray}
where $c=1,\ldots, N_c^2-1$ is the group index, with $N_c=2$ for
$SU(2)$, $f$ is the axion decay constant,  $F_{\mu \nu}^a=
\partial_\mu A_\nu^a - \partial_\nu A_\mu^a +g \epsilon^{abc} A_\mu^b
A_\nu^c$ is the  gauge field tensor, $\tilde{F}^{c\,\mu \nu}$ is its
dual,
\begin{equation}
\tilde{F}^{c\,\mu \nu}=\frac{\epsilon^{\mu \nu \rho \sigma}}{2 a^3(t)}
F_{\rho \sigma}^c,
\end{equation}
and $a(t)$ is the scale factor. 

The $SU(2)$ gauge field admits a homogeneous vacuum expectation value
(VEV), defined
as~\cite{Maleknejad:2011jw,Maleknejad:2016qjz,Adshead:2012kp}
\begin{eqnarray}
 \bar{A}^c_{i}=a(t)\psi(t)\delta^c_{i},\;\;\;\;\bar{A}^c_0=0,
\label{Ansatz}
\end{eqnarray}
which leads to the gauge field strength components,
\begin{eqnarray}
&& \bar{F}^{c\,0i}= -\bar{F}^{c,i0}= -\frac{1}{a} (H\psi + \dot \psi)
  \delta^{ci},\;\;\; \bar{F}^{c\,ij}= \frac{g}{a^2} \psi^2
  \epsilon^{cij}, \nonumber\\ && \bar{F}_{0i}^c= -\bar{F}_{i0}^c=
  a(H\psi + \dot \psi) \delta_i^c,\;\;\; \bar{F}_{ij}^c= g a^2 \psi^2
  \epsilon_{ij}^c.
\label{barF}
\end{eqnarray}

Note that in
chromoinflation~\cite{Maleknejad:2011jw,Maleknejad:2016qjz,Adshead:2012kp}
the coupling $\lambda$ between the inflaton $\phi$ and the gauge
fields in the interaction term in eq.~(\ref{lagr}) is assumed to be an
independent constant.  Typically, a consistent background evolution
for the inflaton and gauge field is achieved for large couplings, or
more generally, $\lambda M_{\rm Pl}/f \gg 1$. How to achieve such
large couplings in effective theories has been discussed recently for
example in refs.~\cite{Agrawal:2018mkd,Holland:2020jdh}.  Note that in
the axion dynamics case, the coupling $\lambda$ is simply related to
the gauge coupling $g$ by $\lambda = g^2/(8 \pi^2)$.

The inflaton (axion) field has been shown to dissipate into radiation
bath gauge fields, which forms a thermal bath with temperature  $T$
and whose dissipation coefficient is given
by~\cite{Berghaus:2019whh,Laine:2021ego} 
\begin{equation}
\Upsilon(T)= \kappa \frac{T^3}{f^2},
\label{Upsilon}
\end{equation}
where $\kappa$ is given by
\begin{equation}
\kappa \simeq 1.2 \pi\frac{g^4 (g^2 N_c)^3
  (N_c^2-1)}{(64\pi^3)^2}\left[\ln\left(\frac{m_D}{\gamma}\right)+3.041\right],
\label{kappa}
\end{equation}
where $m_D^2=g^2 N_c T^2/3$ is the Debye mass squared of the
Yang-Mills plasma and $\gamma$ is given by the solution
of~\cite{Moore:2010jd}
\begin{equation}
\gamma=\frac{g^2 N_c T}{4 \pi}
\left[\ln\left(\frac{m_D}{\gamma}\right)+3.041\right].
\end{equation}
Explicitly, this gives for $\kappa$ the result
\begin{eqnarray}
\kappa &= & 0.3 \alpha_g^5 N_c^3 (N_c^2-1) W\left(
e^{3.041}\sqrt{\frac{4\pi}{3\alpha_g N_c} }\right),
\label{kappa2}
\end{eqnarray}
where $\alpha_g=g^2/(4\pi)$ is the fine structure Yang-Mills coupling and $W(x)$ is the
Lambert function, given by the principal solution of $x=w e^w$. {}For
$N_c=2$ and assuming e.g. $\alpha_g=0.1$, we obtain that $\kappa \simeq
2.4 \times 10^{-4}$.
 
\subsection{Background Evolution}

Implementations of warm inflation in the context of chromoinflation
have been considered in some recent
works~\cite{Yeasmin:2022ncm,Mukuno:2024yoa}.  At the background level,
the energy density related to the inflaton field $\phi$, the VEV
$\psi$ of the Yang-Mills gauge field and the thermal bath are,
respectively, given by
 \begin{eqnarray}
&& \rho_{\phi}=\frac{1}{2}\dot{\phi}^2+V(\phi),
\label{rhophi}
\\ && \rho_{YM}=\frac{3}{2} \left( \dot \psi + H \psi\right)^2 +
\frac{3}{2}g^2 \psi^4,
\label{rhoYM}
\\ && \rho_{r}=C_r T^4,
\label{rhor}
 \end{eqnarray}
where $C_r = g_* \pi^2/30$ and $g_*$ denotes the radiation bath
degrees of freedom.  In the present work we assume that the radiation
bath is constituted primarily by the gauge field fluctuations, which
then gives $g_*=2(N_c^2-1)=6$ for the case of the  (massless) $SU(2)$
Yang-Mills fields\footnote{When including also the inflaton
fluctuations as thermalized, then $g_*=2N_c^2-1$.}.

Likewise, the evolution equations for the inflaton field $\phi$, the
Yang-Mills VEV $\psi$ and the radiation energy density $\rho_r$ are
given by
  \begin{eqnarray}\label{eqphi0}
&&\!\!\!\!\!\!\!\!\!\!\! \!\!\!\!  \ddot{\phi}+3H\dot{\phi}+V_{,\phi}
    +
    3\frac{\lambda}{f}g\psi^2(\dot{\psi}+H\psi)+\Upsilon\dot{\phi}=0,
\\ && \!\!\!\!\!\!\!\!\!\!\! \!\!\!\!
\ddot{\psi}+3H\dot{\psi}+(2H^2+\dot{H})\psi+2g^2\psi^3-\frac{\lambda
  g}{f}\psi^2\dot{\phi}=0,
\label{eqpsi0}
\\ && \!\!\!\!\!\!\!\!\!\!\! \!\!\!\!
\dot{\rho}_{r}+4H\rho_{r}=\Upsilon\dot{\phi}^2,
\label{eqrhor}
  \end{eqnarray}
where the dissipation coefficient $\Upsilon$ is given by
eq.~(\ref{Upsilon}).

The slow-roll variables can be defined as~\cite{Maleknejad:2016qjz}
 \begin{eqnarray}\label{Slow-roll}
   \epsilon_{\psi}=\frac{\dot{\psi}}{H\psi}~~~~~~\eta_{\psi}=-\frac{\ddot{\psi}}{H\psi}
  \\ \nonumber
 \epsilon_{H}=-\frac{\dot{H}}{H^2}~~~~~~~~\eta_{H}=-\frac{\ddot{H}}{2H\dot{H}}
 \end{eqnarray}
In the slow-roll approximation, $\epsilon_{\psi}<1$, $\eta_{\psi}<1$,
$\epsilon_{H}<1 $ and out of the instability
condition~\cite{Adshead:2013nka,Namba:2013kia,Dimastrogiovanni:2012ew}
(where the gauge perturbations become tachyonic),
i.e., $g^2\psi^2/(2H^2)>1$, the evolution of $\psi$ in
eq.~(\ref{eqpsi0}) is simplified to
\begin{eqnarray}
g\psi^3 \simeq \frac{\lambda}{2f}\psi^2 \dot{\phi}.
\end{eqnarray}  

There are several consistency conditions that need to be satisfied
such that the above equations can work properly in a warm inflation
context. {}First, a thermal bath of particles needs to be produced
during inflation. This has been explicitly verified in
refs.~\cite{Laine:2021ego,DeRocco:2021rzv}.  Second, backreaction from
the nonlinearities which are inherent of non-Abelian gauge fields needs
to be sufficient suppressed such that the effective theory leading to
the evolution equations (\ref{eqphi0}), (\ref{eqpsi0}) and
(\ref{eqrhor}) remains valid. This typically requires that we work with
temperatures that remain smaller than the axion decay
constant~\cite{DeRocco:2021rzv}, i.e.,  $T<f$. The warm inflation
regime itself also requires $T>H$. So, we must ensure the hierarchy of
energy scales: $H<T<f$.
Third, to avoid the instabilities of the non-Abelian gauge field, we
are required to have $g|\psi|/H> \sqrt{2}$.  {}Forth, perturbativity
also requires that the gauge coupling to be small, $\alpha_g<1$, such
that we can also trust in the computation leading to the dissipation
coefficient eq.~(\ref{Upsilon}). {}Fifth, to ensure that there is a
thermal bath of gauge particles, the temperature must remain above the
confinement scale of the model~\cite{Berghaus:2019whh}. This implies
in particular in the choice for the inflaton potential $V(\phi)$. The
usual low energy infrared (IR) potential assumed for the axion cannot be used
here and some ultraviolet potential needs to be used. Some different potentials
have been considered before, like an hybrid type of potential as used
in ref.~\cite{Berghaus:2019whh}. An exponential potential was
considered in ref.~\cite{Goswami:2019ehb}. A simple quadratic potential
for the inflaton was used in ref.~\cite{Laine:2021ego}, while in
ref.~\cite{DeRocco:2021rzv} some axion monodromy type of potentials
have been considered.

\section{Thermal contributions to the background dynamics}
\label{sec3}

To take into account the effects of the thermal bath on the background
dynamics, we will work analogously to the loop expansion in quantum field
theory.  This starts by expanding the fields around their vacuum
background expectation values. In this work in particular, we will be
looking how the fluctuations around the gauge field background $\psi$
will affect the dynamics. Hence, the gauge field in (\ref{lagr}) is
expanded like $A_\mu \to A'_\mu = \bar{A}_\mu(t) + A_\mu$, where
$\langle A'_\mu \rangle = \bar{A}_\mu(t)$, with  $\bar{A}_\mu(t)$
given by eq.~(\ref{Ansatz}) and $\langle A_\mu \rangle = 0$.  Hence, 
we have for instance that
\begin{eqnarray}
F^a_{\mu \nu} = \bar{F}_{\mu \nu}^a + \left(\bar{D}_\mu A_\nu
\right)^a - \left(\bar{D}_\nu A_\mu \right)^a + g \epsilon^{abc}
A_\mu^b A_\nu^c,
\label{Fexp}
\end{eqnarray}
where $\bar{F}_{\mu \nu}^a$ is given by eq.~(\ref{barF}) and
\begin{equation}
\bar{D}_\mu  = \partial_\mu -i g J^a \bar{A}_\mu^a,
\end{equation}
is the covariant derivative expressed in terms of the background gauge
field, with $J^a$ the generators of the $SU(2)$ group. Working in the
adjoint representation of the $SU(2)$ gauge group,
$(J_a)_{bc}=\epsilon_{abc}$.

\begin{center}
\begin{figure}[!htb]
\includegraphics[width=6cm]{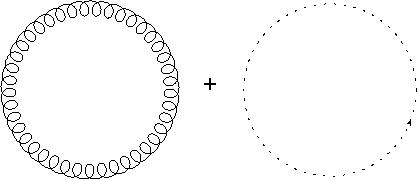}
\caption{The vacuum diagrams contributing at leading order to
the effective potential for $\psi$. The first diagram is a gauge loop diagram, with
propagator with a gauge mass squared $m_A^2=2 g^2 \psi^2$, while the second diagram is the 
ghost contribution, with propagator with mass  $m_\eta=m_A$.}
\label{fig1}
\end{figure}
\end{center}

Using eq.~(\ref{Fexp}) in (\ref{lagr}), the action, up to second order
in the gauge field fluctuations, then takes the form
\begin{equation}
S\simeq S_0 + \delta^2S,
\label{S0S2}
\end{equation}
where 
\begin{eqnarray}
S_0 &=&  \int d^4 x a^3(t) \left[\frac{1}{2}(\partial_\mu \phi)^2 -
  V(\phi) -\frac{1}{4}\bar{F}_{\mu\nu}^c \bar{F}^{c \,\mu \nu} \right.
  \nonumber \\ &-& \left. \frac{\lambda}{4f}\phi \frac{\epsilon^{\mu
      \nu \rho \sigma}}{2 a^3(t)}  \bar{F}_{\mu \nu}^c \bar{F}_{\rho
    \sigma}^c\right] \nonumber \\ &=& \int d^4 x\, a^3 \left[
  \frac{\dot \phi^2}{2}-V(\phi) +\frac{3}{2} \left( \dot \psi + H
  \psi\right)^2   \right.  \nonumber \\ &-&
  \left. \frac{3}{2}g^2\psi^4 - 3 \frac{\lambda}{f} g \phi \left( \dot
  \psi + H \psi\right)\psi^2 \right],
\label{S0}
\end{eqnarray}
is the zeroth order action in the fluctuations and $\delta^2S$ is the
action when expanding the fields up to the second-order in the
fluctuations (see also ref.~\cite{Adshead:2013nka} for a similar
approach for deriving the perturbation equations in chromoinflation).
The Yang-Mills and Chern-Simons contributions can then be expressed as
\begin{eqnarray}
\delta^2 S_{\rm YM}  &=& \int d^4 x \, a^3 \left\{ - \frac{1}{2}
(\bar{D}^\mu A^\nu)^a (\bar{D}_\mu A_\nu)^a  \right.  \nonumber \\ &+&
\left. \frac{1}{2} (\bar{D}^\mu A^\nu)^a (\bar{D}_\nu A_\mu)^a
-\frac{1}{2} g \epsilon^{abc} \bar{F}^{a\, \mu \nu} A_\mu^b A_\nu^c
\right.  \nonumber \\ &- & \left. \frac{\lambda}{2f}\phi
\;\frac{\epsilon^{\mu \nu \rho \sigma}}{a^3} \left[ (\bar{D}_\mu
  A_\nu)^a (\bar{D}_\rho A_\sigma)^a  \right.\right.  \nonumber \\ &+&
  \left.\left. \frac{g}{2} \epsilon^{abc} \bar{F}^a_{ \mu \nu}
  A_\rho^b A_\sigma^c \right]\right\}.  \nonumber \\
\label{S2YM}
\end{eqnarray}
We still need to choose a gauge fixing term. Here, it becomes
convenient to fix a gauge that depends explicitly on the background
gauge field, such that (see, e.g.~\cite{Srednicki:2007qs})
\begin{equation}
\delta^2S_{\rm gf}= \int d^4 x\, a^3 \left[ -\frac{1}{2 \alpha}
  (\bar{D}^\mu A^\mu)^a (\bar{D}_\nu A_\nu)^a \right],
\label{gauge}
\end{equation}
and with the corresponding {}Faddeev-Popov ghost term,
\begin{equation}
\delta^2S_{\rm ghost}= \int d^4 x\, a^3 \left[  (\bar{D}^\mu
  \bar{\eta})^a (\bar{D}_\mu \eta)^a \right],
\label{ghost}
\end{equation}
where $\eta$ and $\bar{\eta}$ are the Faddeev-Popov ghost fields.
Hence, by choosing the gauge $\alpha=1$ and after some manipulation of
indexes in eq.~(\ref{S2YM}) and integration by parts, we obtain the
result for the quadratic term in the fluctuations in the action (note
that in arriving at the expression below, a term coming from the
integration by parts, but not depending on the background gauge field,
was neglected) 
\begin{eqnarray}
\delta^2 S &=& \int d^4 x \, a^3 \left\{ - \frac{1}{2} (\bar{D}^\mu
A^\nu)^a (\bar{D}_\mu A_\nu)^a  \right.  \nonumber \\ &+&
\left. (\bar{D}^\mu \bar{\eta})^a (\bar{D}_\mu \eta)^a -  g
\epsilon^{abc} \bar{F}^{a\, \mu \nu} A_\mu^b A_\nu^c  \right.
\nonumber \\ &- & \left. \frac{\lambda}{2f}\phi \;\frac{\epsilon^{\mu
    \nu \rho \sigma}}{a^3} \left[ (\bar{D}_\mu A_\nu)^a (\bar{D}_\rho
  A_\sigma)^a  \right.\right.  \nonumber \\ &+&
  \left.\left. \frac{g}{2} \epsilon^{abc} \bar{F}^a_{ \mu \nu}
  A_\rho^b A_\sigma^c \right]\right\}.  \nonumber \\
\label{S2YM2}
\end{eqnarray}
By substituting in eq.~(\ref{S2YM2}) the equations (\ref{Ansatz}) and
(\ref{barF}), we have many terms depending on the background field $\psi$.
Many of them vanish when performing the functional integration over
the fluctuations (e.g. the Chern-Simons dependent terms which are
total derivatives).  {}For the terms effectively contributing, one
notices that in the presence of a nonvanishing background gauge field,
both the gauge field and the ghosts acquire a mass square,
$m_A^2=m_\eta^2=2 g^2 \psi^2$.  Their functional integration lead to
an effective potential for $\psi$ which can be expressed
as\footnote{Note that curvature effects are expected to be negligible
for masses larger than the Hubble scale, e.g. for $2 g^2 \psi^ 2 >
H^2$ and, furthermore, at finite temperature, the relevant momentum
modes for the gauge fluctuations contributing for the loop integrals
are hard momenta, $k \sim T$. In particular, this later condition is well 
satisfied in warm inflation by definition, where $T>H$. This justifies a 
Minskowski like derivation of
the loop terms used to derive the result given by eq.~(\ref{Veff}).}
\begin{eqnarray}
\!\!\!\!\! \Delta V_{\rm eff} (\psi) &=& (N_c^2 -1) \int \frac{d^4
  k_E}{(2 \pi)^4} \ln \left(k_E^2+2 g^2 \psi^ 2\right),
\label{Veff}
\end{eqnarray}
where $k_E$ refers to the (physical) momentum expressed in Euclidean
spacetime.  Diagrammatically, the terms contributing to the effective potential
at leading order are displayed in {}figure~\ref{fig1}.

 {}From the usual rules of finite temperature quantum field
theory~\cite{Kapusta:2006pm}, $k_E^2= k_4^2 + {\bf k}^2$ and
$k_4\equiv \omega_n= 2\pi n T$, $n\in \mathbb{Z}$ and $\omega_n$ are
the Matsubara's frequencies for bosons.  The integrals over momenta
are expressed as
\begin{equation}
\int \frac{d^4 k_E}{(2 \pi)^4} \to T\sum_{k_{4}=\omega_n}\int
\frac{d^{3}k}{\left( 2\pi \right)^{3}}.
\label{sumT}
\end{equation} 
At zero temperature, the momentum integral in eq.~(\ref{Veff}) is
divergent.  The divergent terms can be handled as usual in
perturbation theory in quantum field theory by adding the appropriate
renormalization counterterms in the tree-level action
eq.~(\ref{S0}). The relevant contributions for us are the finite
temperature ones, which, after performing the Matsubara's frequency
sum in eq.~(\ref{Veff}), we are left with the thermal contribution,
\begin{eqnarray}
\Delta V_{\rm eff} (\psi,T) &=& 2(N_c^2 -1) \frac{T^4}{2 \pi^2}J_B(y),
\label{VeffT}
\end{eqnarray}
where~\cite{Kapusta:2006pm}
\begin{eqnarray}
J_B(y) &=& \int_0^{\infty} dx \, x^2 \ln\left( 1- e^{ -\sqrt{x^2+y^2}}
\right) \nonumber \\ &\simeq& -\frac{\pi^4}{45} + \frac{\pi^2}{12}y^2
+ {\cal O}(y^3),
\label{JBy}
\end{eqnarray}
with $y^2= 2 g^2 \psi^2/T^2$ and the last line in eq.~(\ref{JBy})
follows by  using an expansion for $y \ll 1$. We could as well consider the 
additional terms in the expansion in eq.~(\ref{JBy}). However, for our purposes
of demonstrating the effects of the thermal contributions, the truncation of
the series up to quadratic order for the thermal integral $J_B(y)$ suffices 
for us\footnote{We have nevertheless explicitly verified in our numerical analysis
that the higher order terms remain subdominant compared to the quadratic term in 
eq.~(\ref{JBy}).}. We then obtain for
eq.~(\ref{Veff}) the result
\begin{eqnarray}
\Delta V_{\rm eff} (\psi,T) &\simeq& -2(N_c^2 -1) \frac{\pi^2 }{90}
T^4 + (N_c^2-1) g^2 \frac{T^2}{6} \psi^2.  \nonumber \\
\label{VeffT2}
\end{eqnarray}
The total energy density is then modified by the thermal effects, such
that now we have that
\begin{eqnarray}
\rho_T &=& \frac{\dot \phi^2}{2} + V(\phi)  \nonumber \\ &+&
\frac{3}{2} (\dot \psi + H \psi)^2 + \frac{3}{2} g^2 \psi^4  \nonumber
\\ &+& \Delta V_{\rm eff} (\psi,T)+ T s,
\label{rhoT}
\end{eqnarray}
where we opted from now on to work for convenience with the entropy
density $s$,
\begin{eqnarray}
s&=& -\frac{\partial \Delta V_{\rm eff}}{\partial T}   \simeq \frac{4
  \pi^2 (N_c^2-1) T^3}{45} - \frac{(N_c^2-1) g^2 T \psi^2}{3}.
\nonumber \\
\label{entropy}
\end{eqnarray}
We thus see that the presence of the thermal bath in warm chromoinflation 
produces not only a friction term due to the nonperturbative
spharelon effects in the gauge vacua, but also leads to a thermal
plasma mass for the background gauge field $\psi$,
\begin{equation}
m_\psi^2(T) = (N_c^2-1) \frac{g^2 T^2}{3}.
\label{mpsiT}
\end{equation}

The background field equations in warm chromoinflation then
get modified such that now we have that
\begin{eqnarray}
&& \!\!\!\!\!\!\!\!\!\!\! \!\!\!\!
  \ddot{\phi}+3H\dot{\phi}+V_{,\phi} +
  3\frac{\lambda}{f}g\psi^2(\dot{\psi}+H\psi) +\Upsilon\dot{\phi}=0,
\label{eqphi}
\\ && \!\!\!\!\!\!\!\!\!\!\! \!\!\!\!
\ddot{\psi}+3H\dot{\psi}+(2H^2+\dot{H})\psi+2g^2\psi^3  \nonumber
\\ && + m_\psi^2(T) \psi  -\frac{\lambda g}{f}\psi^2\dot{\phi}=0,
\label{eqpsi}
\\ && \!\!\!\!\!\!\!\!\!\!\! \!\!\!\!  T\dot{s}+3H T s
=\Upsilon\dot{\phi}^2,
\label{eqs}
\end{eqnarray}
and with the Hubble parameter given by
\begin{eqnarray}
H^2 &=& \frac{1}{3 M_{\rm Pl}^2} \left[\frac{\dot \phi^2}{2} + V(\phi)
  +\frac{3}{2} (\dot \psi + H \psi)^2 \right. \nonumber \\ &+& \left.
  \frac{3}{2} g^2 \psi^4  + \Delta V_{\rm eff} (\psi,T)+ T s \right].
\label{hubble}
\end{eqnarray}

We can qualitatively already investigate the effect of the thermal mass term
in the background equation for $\psi$.  In the slow-roll
approximation, from eq.~(\ref{eqpsi}), we obtain that
\begin{eqnarray}
3H\dot{\psi} \simeq - \left[2H^2+m_\psi^2(T) \right] \psi- 2g^2\psi^3
+\frac{\lambda g}{f}\psi^2\dot{\phi}.
\label{slowpsi}
\end{eqnarray}
If we associate the right-hand side in eq.~(\ref{slowpsi}) with an
analogous of the field derivative of an effective potential for $\psi$
(see, e.g.  ref.~\cite{Fujita:2021eue} for a similar approach in cold
chromoinflation), then we can define the effective potential (and
assuming $\dot \phi$ as constant),
\begin{equation}
V_{\rm eff}(\psi, T) \simeq \left[2H^2+m_\psi^2(T) \right]
\frac{\psi^2}{2} -\frac{\lambda g \dot{\phi}}{f} \frac{\psi^3}{3}+
g^2\frac{\psi^4}{2}.
\label{VeffpsiT} 
\end{equation}
By defining the dimensionless quantities,
\begin{eqnarray}
\Psi &=& \frac{g \psi}{H} , \nonumber\\ \xi &=& \frac{\lambda \dot
  \phi}{2 f H},
\label{definitions}
\\ M_\psi  &=& \frac{m_\psi(T)}{H}, \nonumber
\end{eqnarray}
we are left with
\begin{equation}
\bar{V}_{\rm eff}(\Psi) = \left(2+M_\psi^2\right) \frac{\Psi^2}{2}
-\frac{2 \xi}{3 }\Psi^3+ \frac{\Psi^4}{2},
\label{VeffPsi} 
\end{equation}
where $\bar{V}_{\rm eff}(\Psi) = g^2V_{\rm eff}(\psi, T) /H^4$.  The
potential $\bar{V}_{\rm eff}(\Psi) $ has in general three extrema:  at
$\Psi=0$ and at
\begin{equation}
\Psi_\pm = \frac{1}{2} \left[ \xi \pm \sqrt{ \xi^2 -
    2\left(2+M_\psi^2\right)}\right].
\label{Psipm}
\end{equation}
It can be easily checked that $\Psi=0$ is always a minimum of
$\bar{V}_{\rm eff}(\Psi)$, while $\Psi_-$ is a maximum and $\Psi_+$ is
another minimum provided that $\xi^2 > 2\left(2+M_\psi^2\right)$. At
$\xi^2 = 2\left(2+M_\psi^2\right)$,  $\Psi_\pm$ coalesces to an
inflection point of $\bar{V}_{\rm eff}(\Psi)$ located at
$\Psi=\xi/2$. {}For $\xi^2 < 2\left(2+M_\psi^2\right)$, we have that
$\Psi=0$ is the only solution. {}Finally, there is a value for $\xi$,
where the minimum at the origin is degenerate with the minimum at
$\Psi_+$, i.e., $\bar{V}_{\rm eff}(0)=\bar{V}_{\rm eff}(\Psi_+)$,
which happens at
\begin{equation}
\xi^2=\frac{9}{4} \left(2+M_\psi^2\right).
\label{xic}
\end{equation}
This behavior of the effective potential, as seen here as function of the values
of the parameters, resembles exactly that one for a first-order phase
transition as a function of the temperature. The value of the
temperature for which $\bar{V}_{\rm eff}(0)=\bar{V}_{\rm eff}(\Psi_+)$
is satisfied would correspond to the critical temperature of the phase
transition, as conventionally considered in the literature of
Ginzburg-Landau phase transition when it is first order~\cite{Goldenfeld:1992qy}.  
We can here
estimate this critical temperature if we use the slow-roll
approximation obtained from the entropy evolution equation,
eq.~(\ref{eqs}),
\begin{equation}
Ts \simeq \frac{\Upsilon}{3H} \dot\phi^2 = \frac{\kappa T^3}{ 3fH}
\dot\phi^2,
\label{Ts}
\end{equation}
where we have used the expression (\ref{Upsilon}) for the dissipation
coefficient.  {}Finally, from eq.~(\ref{entropy}), considering $g \psi
\ll T$, and from the definitions given in eq.~(\ref{definitions}), we obtain
the result for the critical temperature for phase transition in terms
of the parameters of the model,
\begin{eqnarray}
\frac{T_c}{H} &\sim &  \frac{2 \pi^2 \lambda^2}{45 g^2\kappa } \left[
  1 + \sqrt{  1 - \frac{6075 g^2 \kappa^2}{2(N_c^2-1) \pi^4 \lambda^4}
  } \,\right] \sim  \frac{4 \pi^2 \lambda^2}{45 g^2\kappa }.
\nonumber \\
\label{Tc}
\end{eqnarray}

In warm inflation we have in general that $T/H$ increases during
inflation~\cite{Motaharfar:2021egj,Das:2022ubr}.  Thus, after the
critical temperature eq.~(\ref{Tc}) is reached, $\Psi_+$ becomes a
local minimum. Hence, above $T_c$ it becomes energetically favored for
the system to be driven to $\psi=0$. This happens rather fast, in less
than one e-fold, as indicated by our numerical results, after which the 
gauge vacuum expectation value no
longer plays a role in the dynamics.  The above qualitative analytical
analysis is confirmed by our numerical results that are presented in
the next section.  

\section{Numerical Results}
\label{sec4}

\begin{center}
\begin{figure*}[!bth]
\subfigure[]{\includegraphics[width=7.cm]{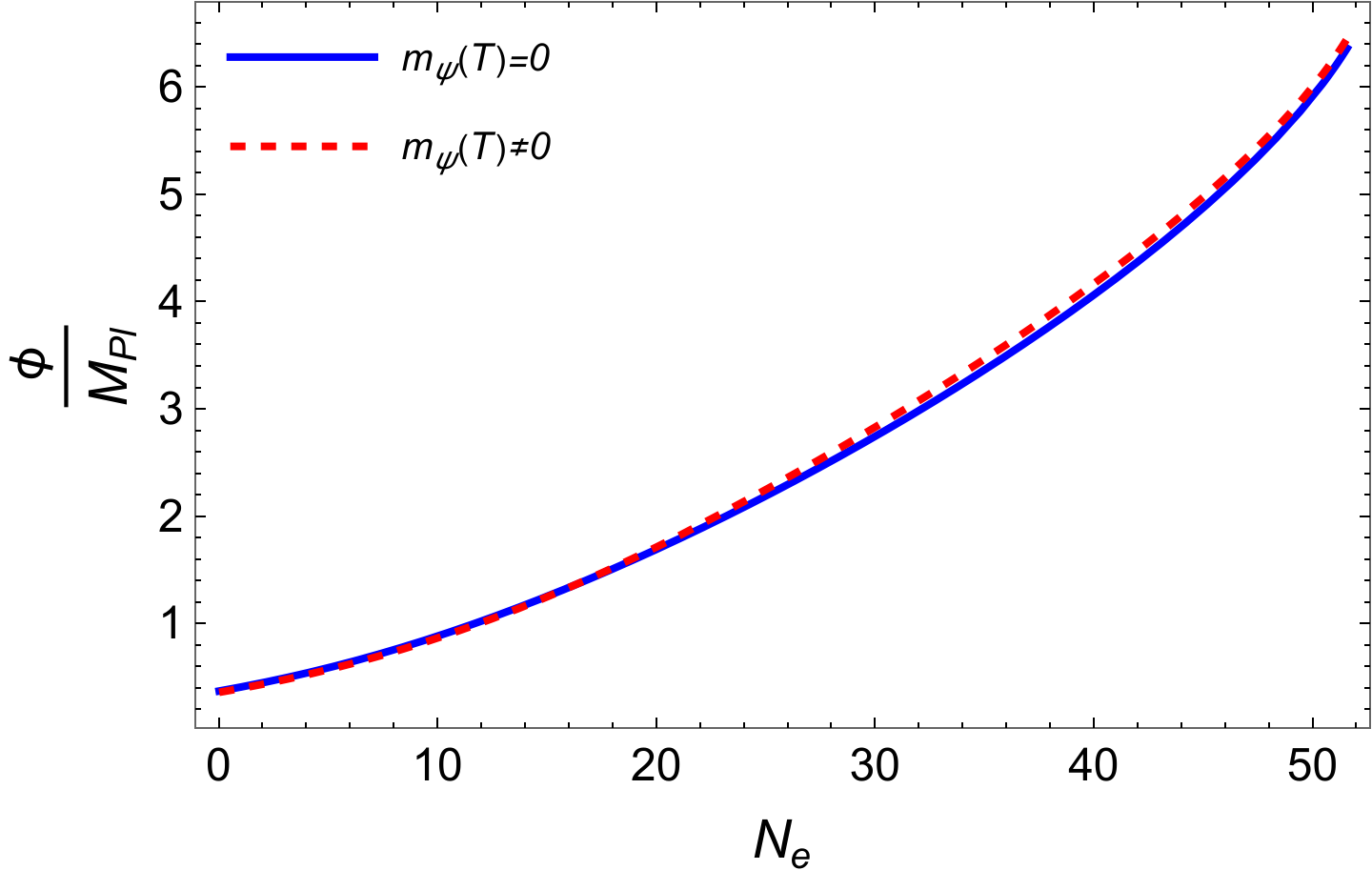}}
\subfigure[]{\includegraphics[width=7.5cm]{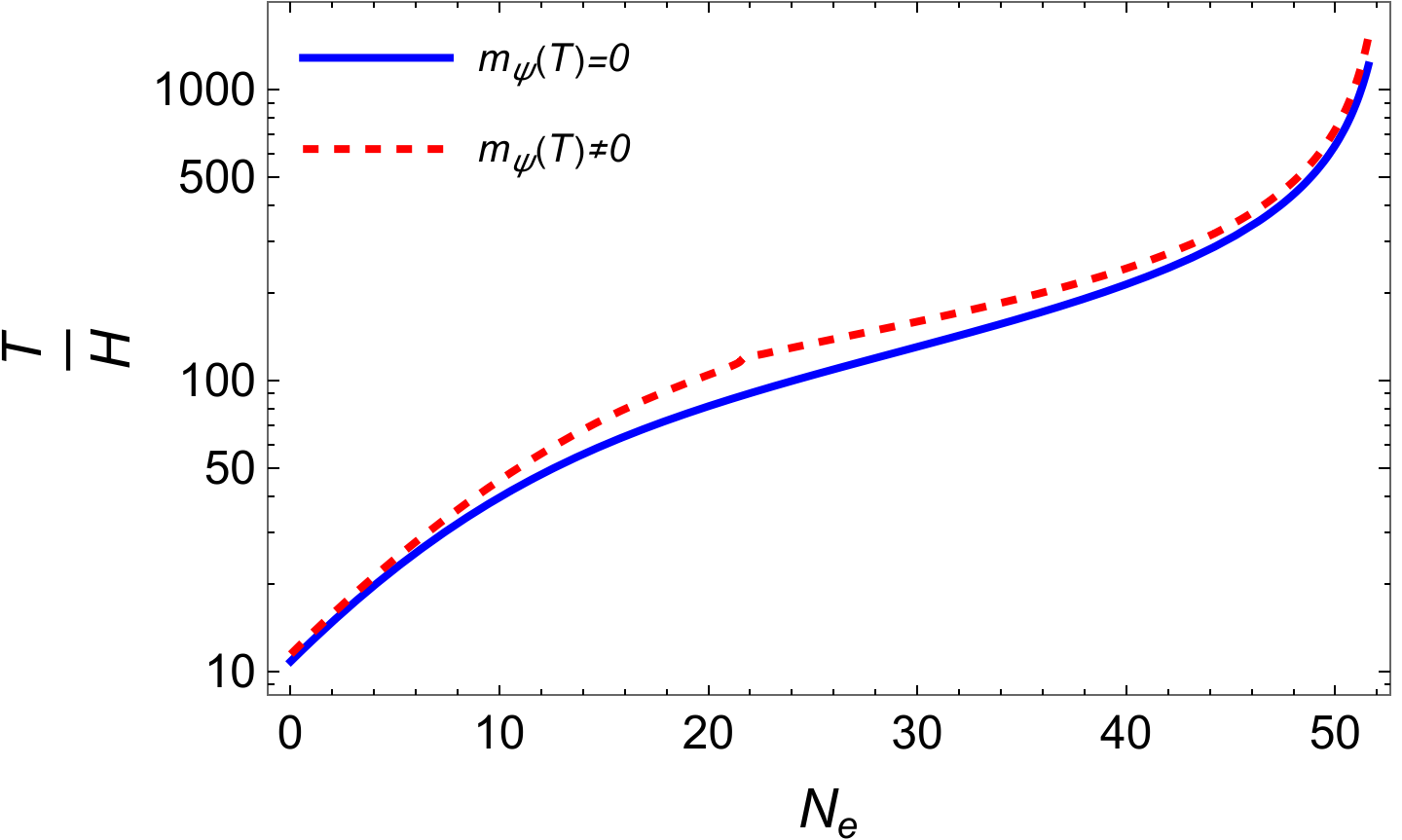}}
\subfigure[]{\includegraphics[width=7.cm]{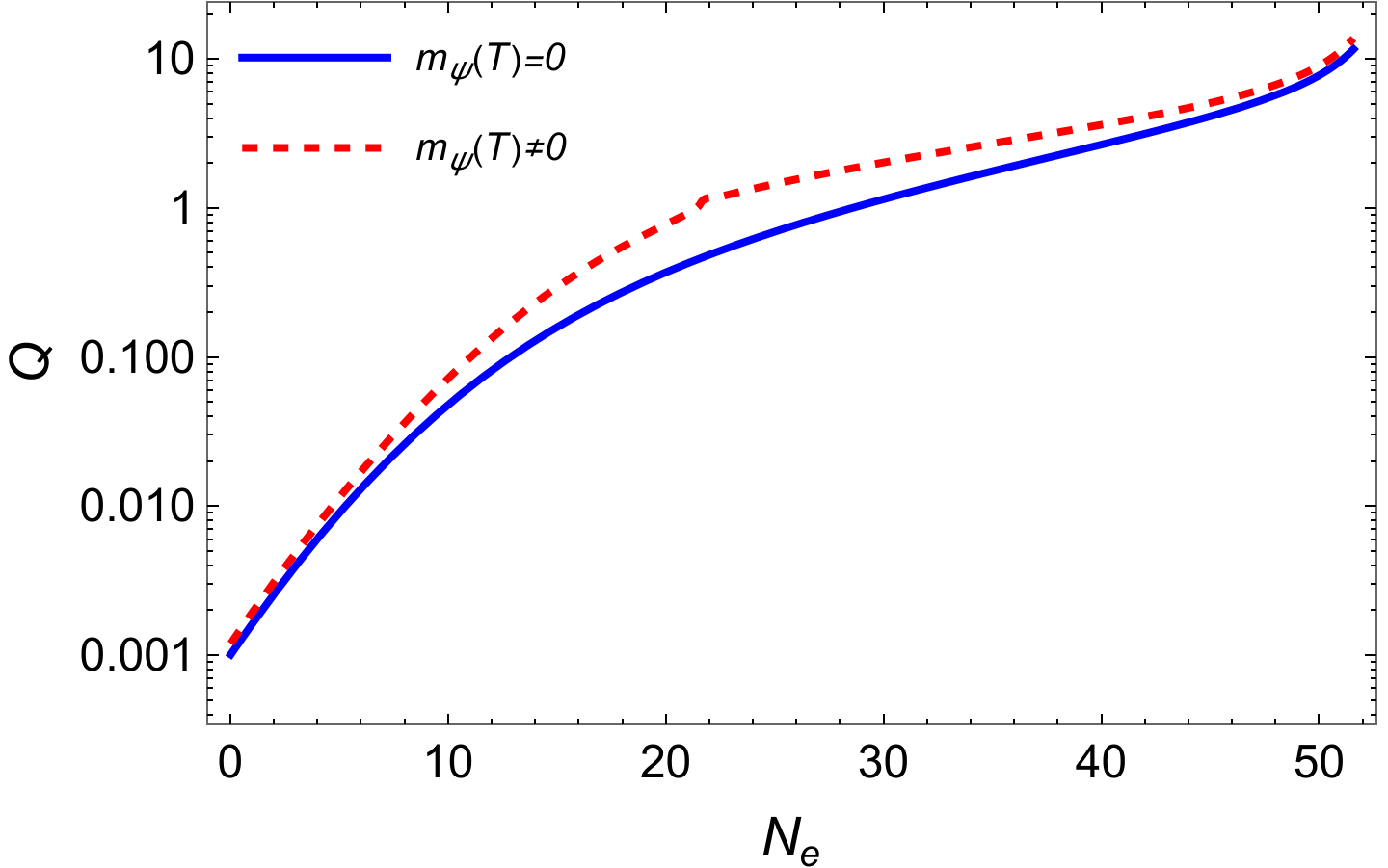}}
\subfigure[]{\includegraphics[width=7.5cm]{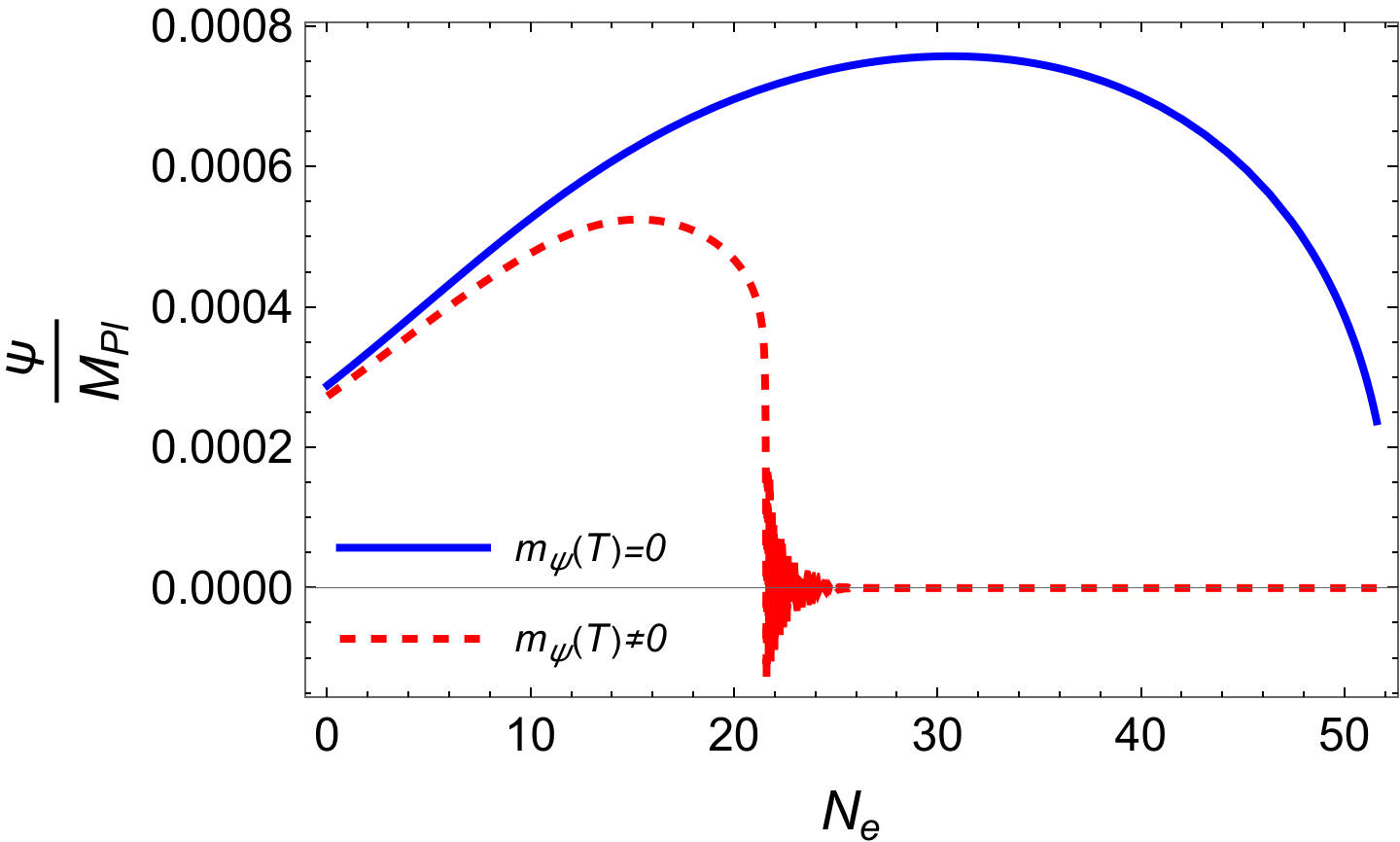}}
\caption{The inflaton field (panel a), temperature over Hubble
  parameter (panel b), the dissipation ratio $Q=\Upsilon/(3H)$ (panel
  c) and the gauge field background (panel d) as a function of the
  number of e-folds. The initial value of the dissipation ratio is
  $Q_0=1.2\times 10^{-3}$.  The solid lines correspond to the results obtained
in the case where the
  thermal contribution to the gauge background is neglected, while
  the dashed lines give the results when it is included.}
\label{fig2}
\end{figure*}
\end{center}

\begin{center}
\begin{figure*}[!htb]
\subfigure[]{\includegraphics[width=7.cm]{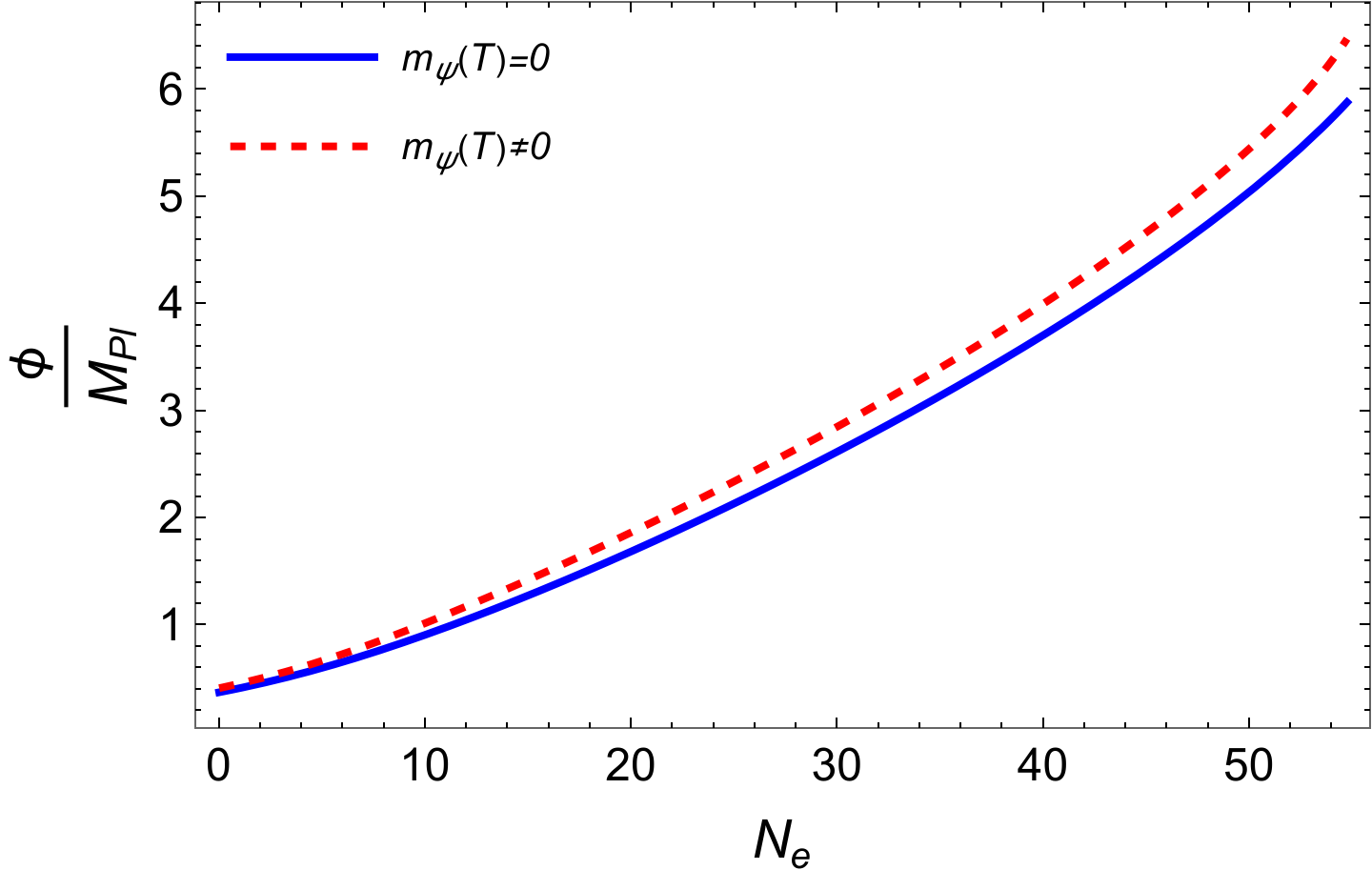}}
\subfigure[]{\includegraphics[width=7.5cm]{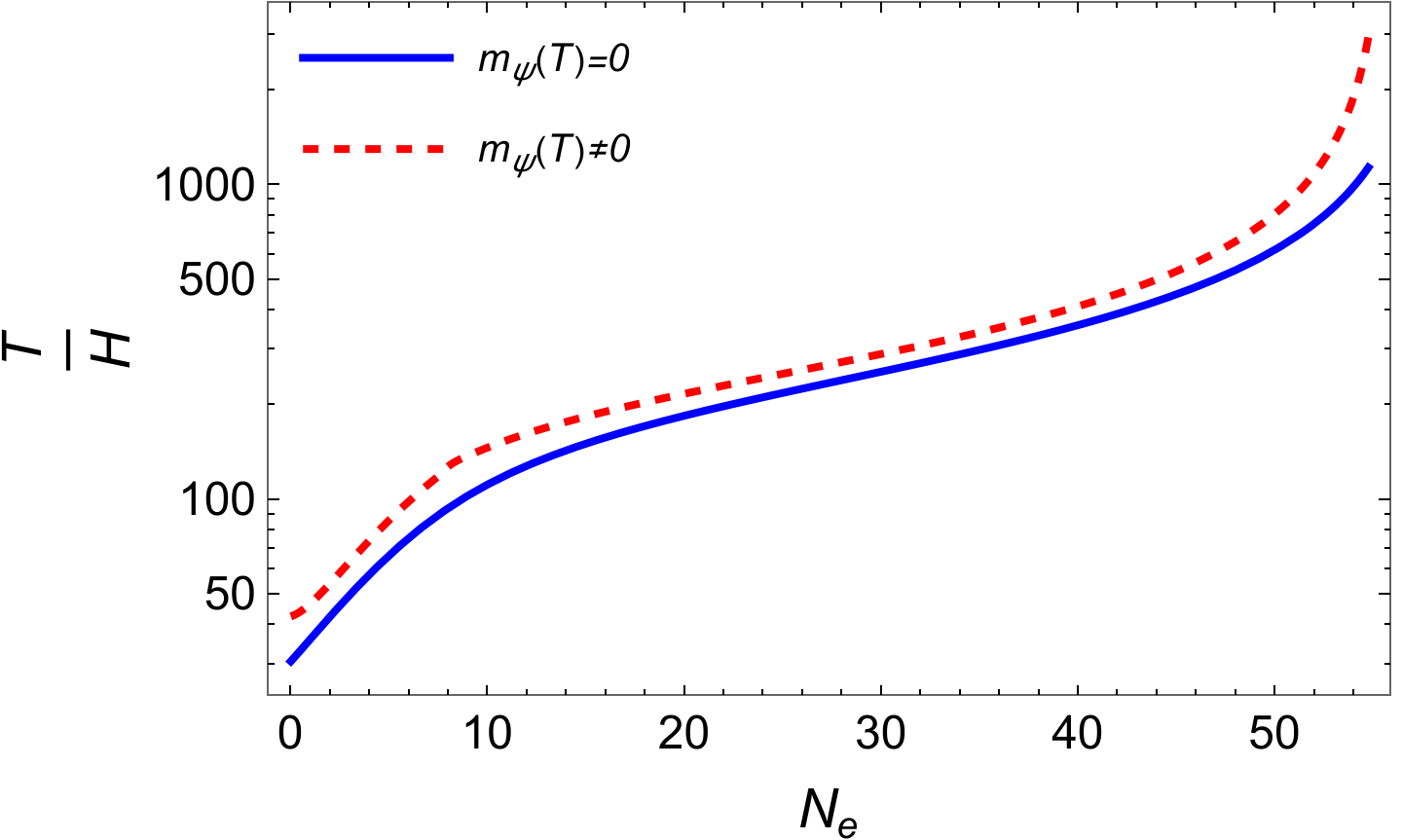}}
\subfigure[]{\includegraphics[width=6.8cm]{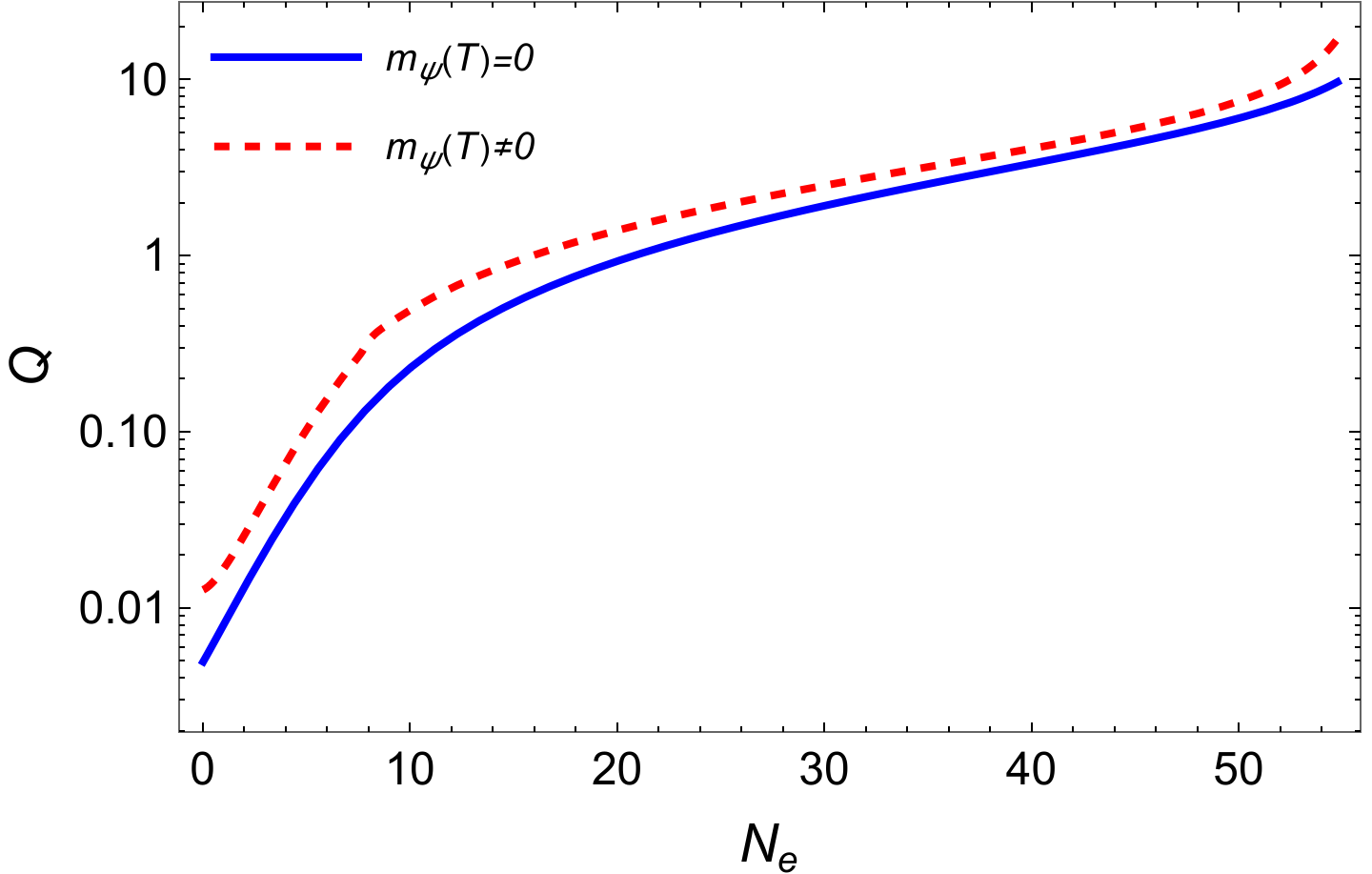}}
\subfigure[]{\includegraphics[width=7.5cm]{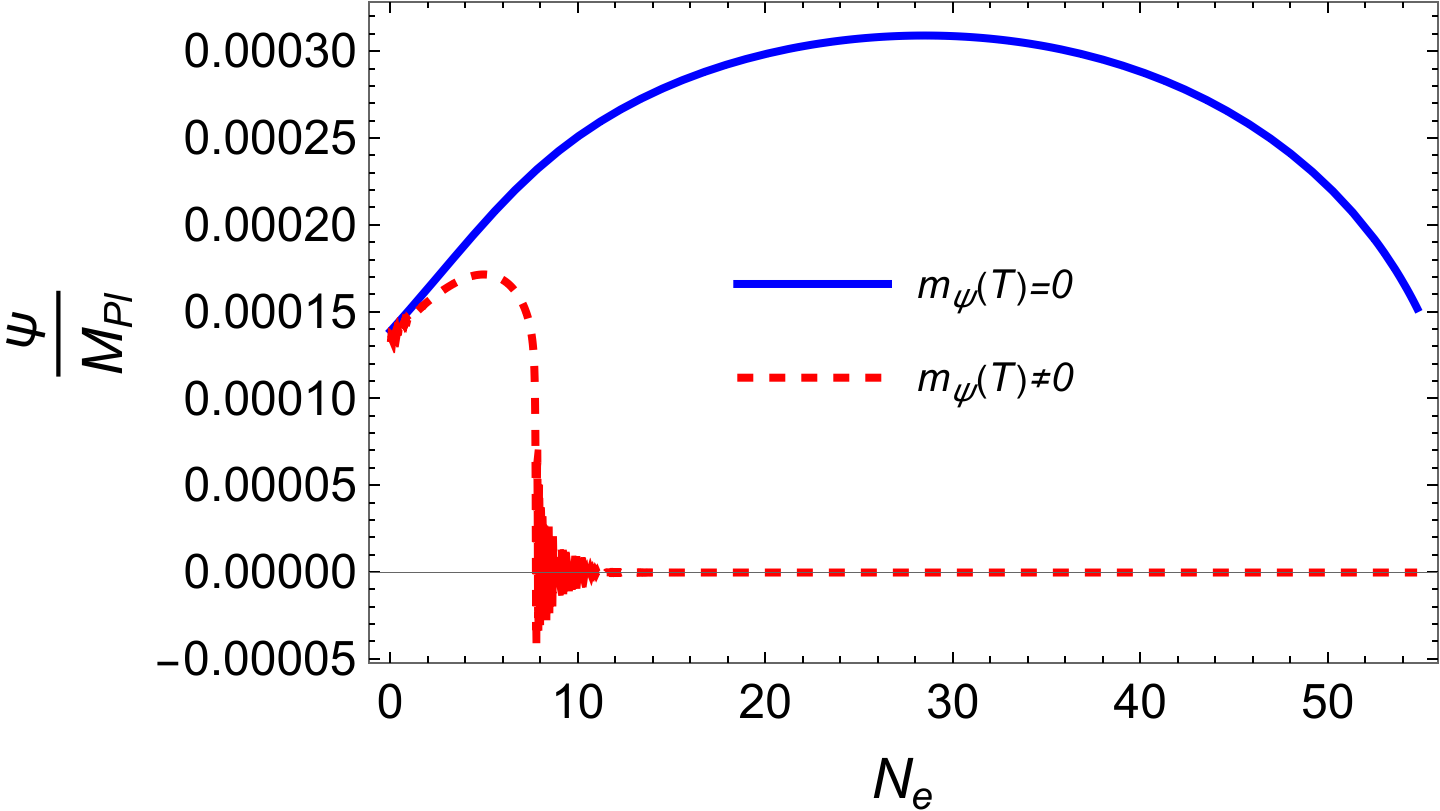}}
\caption{The same as in {}figure~\ref{fig2}, but for an initial value of
  the dissipation ratio is $Q_0=1.3\times 10^{-2}$.}
\label{fig3}
\end{figure*}
\end{center}

To illustrate our results, let us consider an axion-like potential for
the inflaton and given by
\begin{equation}
V(\phi) = \Lambda^4 \left[1+ \cos(\phi/f_D)\right],
\label{pot}
\end{equation}
where $\Lambda$ is the normalization of the potential (typically
$\Lambda^4 = m_\phi^2 f_D^2$, where $m_\phi$ is the mass of the
axion-like particle and $f_D$ its decay constant). Note that in
standard cold natural inflation one typically requires $f_D \gtrsim
M_{\rm Pl}/\sqrt{2}$ to have a slowly rolling regime for the
inflaton~\cite{Freese:2004un}. 
To be concrete, we assume $\Lambda$ to correspond to some confinement
energy scale of some UV complete theory. Inflation here would then
happen at energy scales below $\Lambda$ but still larger than some
IR energy scale for the confinement of the $SU(2)$ gauge field 
considered here, such that the presence of a thermal bath and the 
use of the dissipation coefficient eq.~(\ref{Upsilon}) are justified.

In warm inflation with a dissipation term like eq.~(\ref{Upsilon}) it
was also been shown~\cite{Montefalcone:2022jfw} that consistency of
the natural inflation potential with the  observations still requires
$f_D \sim M_{\rm Pl}$. In particular, a more recent
analysis~\cite{Zell:2024vfn} has also shown that this model can only
sustain inflation consistent with the observations in the weak regime
of warm inflation, in which case $Q=\Upsilon/(3H) \ll 1$. Motivated by
these previous references, here we will
not assume that $f$, which appears in the dissipation and
chromoinflation expressions and that comes from the Chern-Simons interaction, 
and $f_D$, which appears in the potential
are the same\footnote{This can also be interpreted in terms of an effective
value for the coupling $\lambda$, such that it would be analogous to consider 
a modified $\bar \lambda$ given by $\bar \lambda = \lambda f_D/f$.}.  
This will give us some more freedom when playing with the
parameters of the model\footnote{It can be assumed in particular that
$f_D$ is a much larger scale than $f$, which is motivated from recent
works based on clockwork
models~\cite{Kaplan:2015fuy,Choi:2015fiu}.}. By fixing $f_D$,
$\lambda$ and $g$, we can obtain for instance $f$ by demanding to
inflation to last a specific number of efolds for a given dissipation
ratio $Q$. 

\begin{center}
\begin{figure*}[!htb]
\subfigure[]{\includegraphics[width=7.cm]{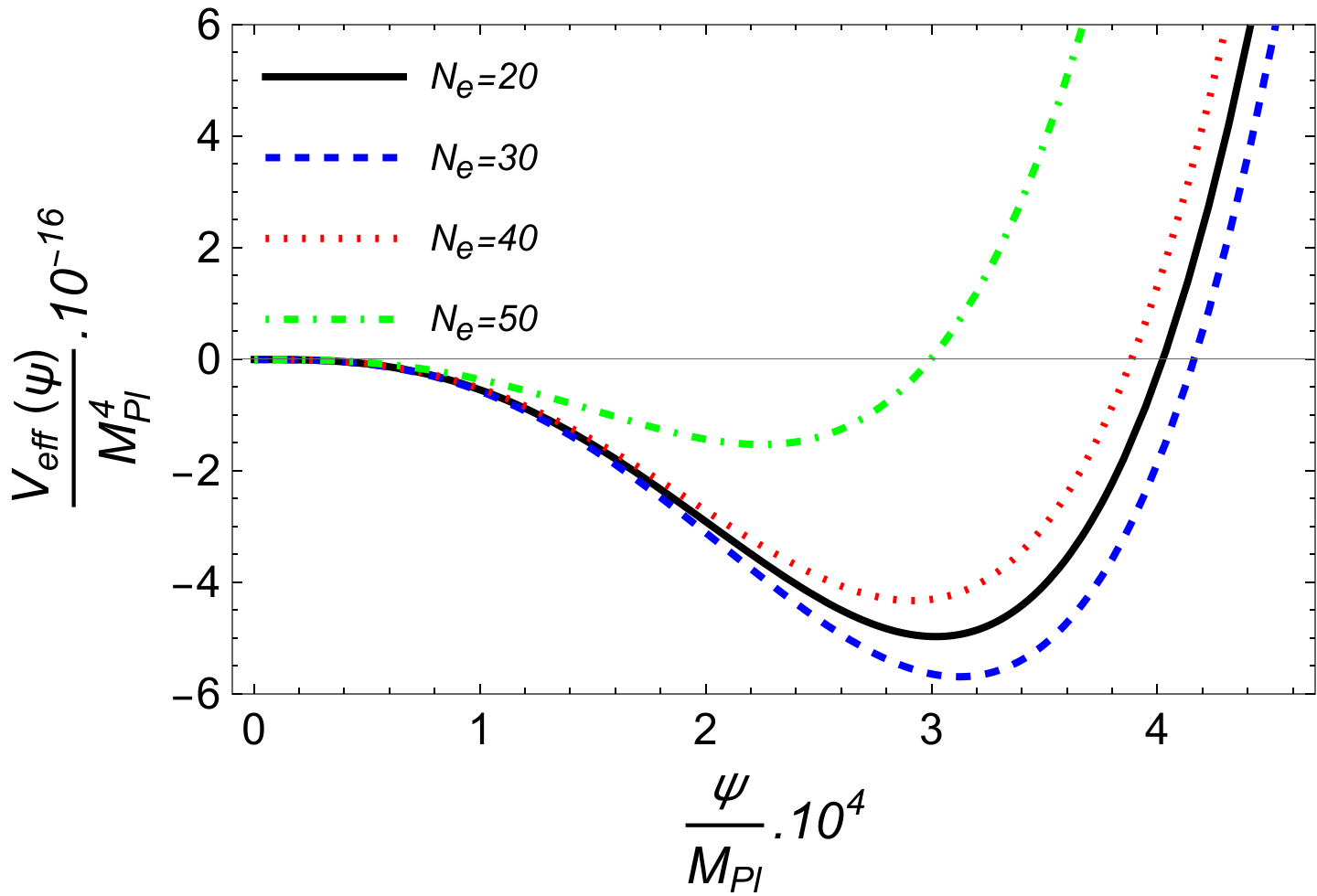}}
\subfigure[]{\includegraphics[width=7.cm]{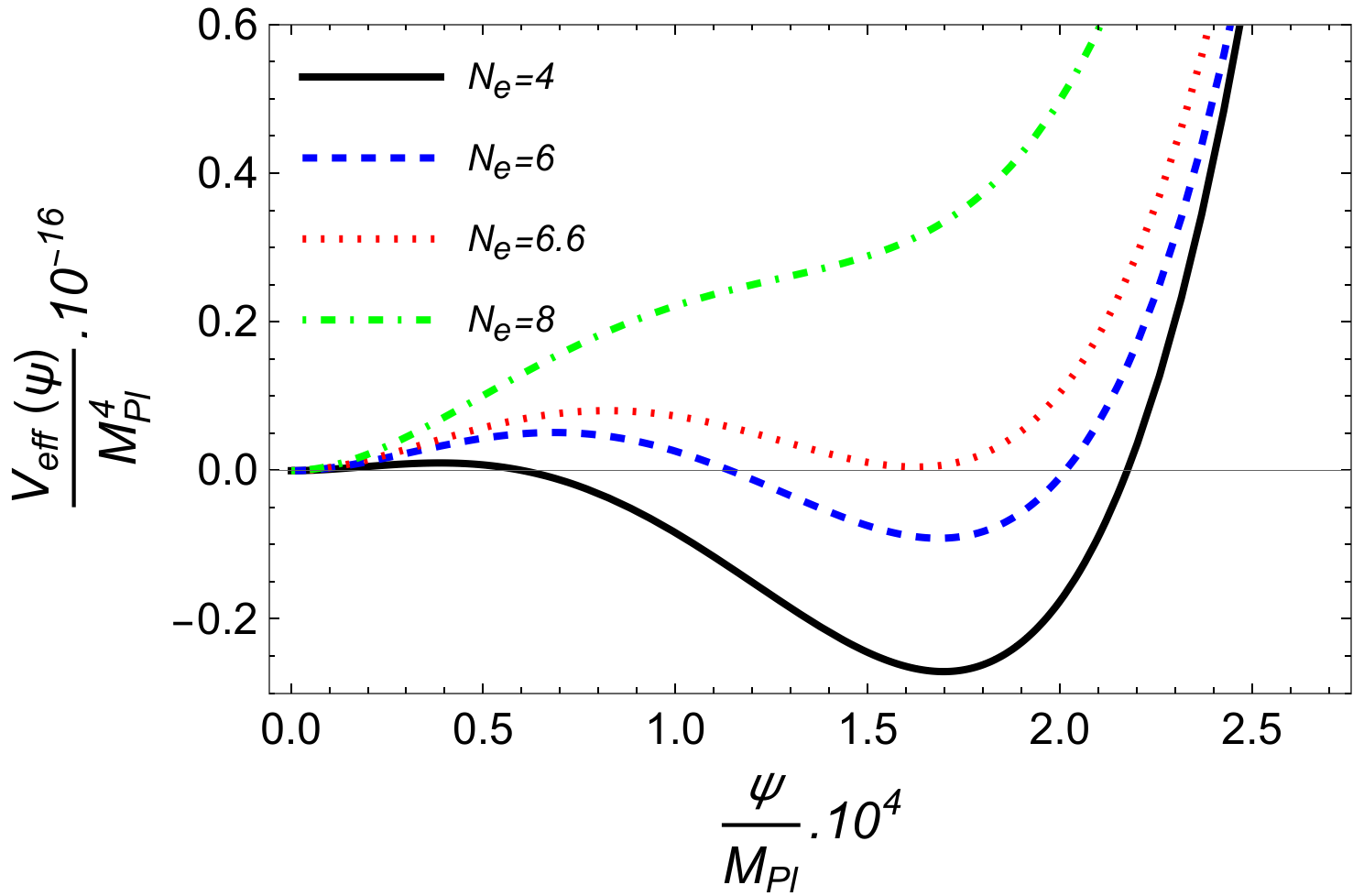}}
\caption{Snapshots of the effective potential eq.~(\ref{VeffpsiT}) as
  a function of $\psi$ at specific values of e-folds. Panel (a) is in
  the absence of the thermal effects for $\psi$, while panel (b) shows
  the effects of including them.}
\label{fig4}
\end{figure*}
\end{center}
\begin{center}
\begin{figure*}[!htb]
\subfigure[]{\includegraphics[width=6cm]{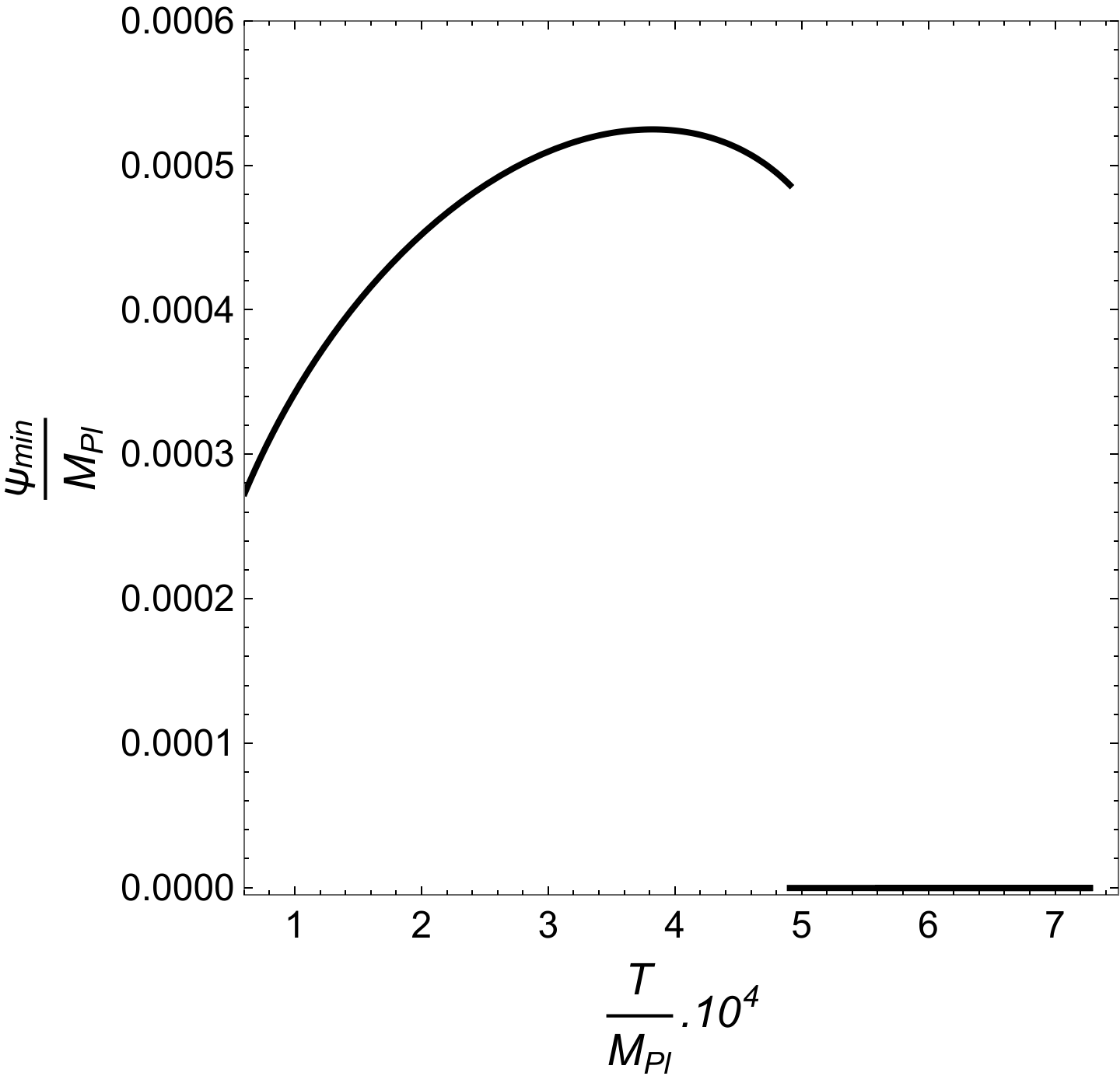}}
\subfigure[]{\includegraphics[width=6cm]{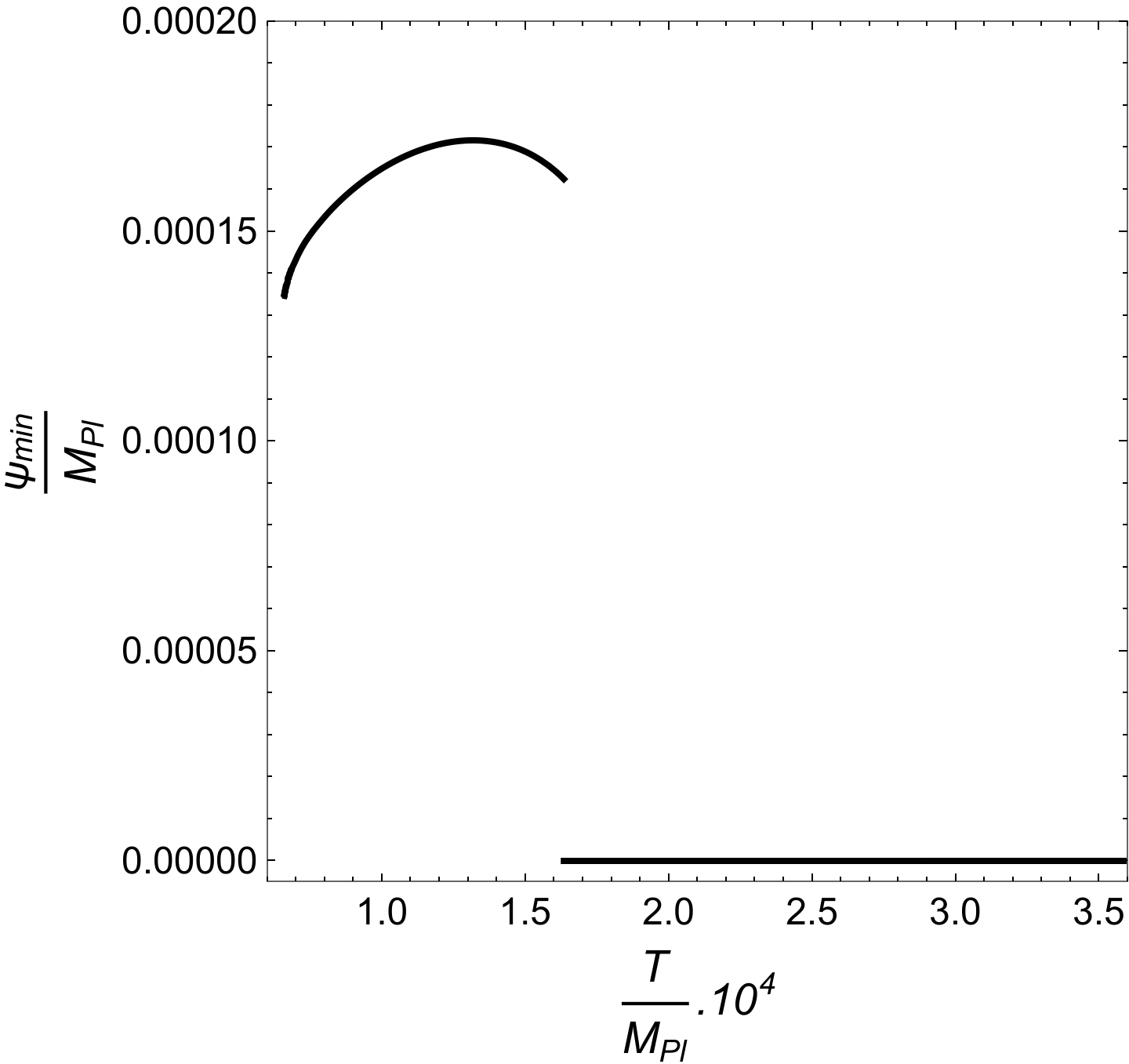}}
\caption{The global minimum of the effective potential
  eq.~(\ref{VeffpsiT}) as a function of the temperature,  for the
  cases of $Q_0=1.2\times 10^{-3}$ (panel a) and for $Q_0=1.3\times
  10^{-2}$ (panel b).}
\label{fig5}
\end{figure*}
\end{center}

Specific examples are given in {}figures~\ref{fig2} and \ref{fig3}, where
we show some of the relevant background quantities. In both examples
shown in those figures the total number of e-folds of inflation was
taken to be $N_{\rm infl}=55$ and for illustration purposes the
parameters taken were $g=0.6$, $\lambda=100$, $f_D/M_{\rm Pl}=
3/\sqrt{2}$ (as motivated e.g. from ref.~\cite{Montefalcone:2022jfw}).  
The initial dissipation ratio in {}figure~\ref{fig2} is
$Q_0=1.2\times 10^{-3}$, from which the values of $f$ and $\Lambda$
are found to be $f\simeq 0.05 M_{\rm Pl}$ and $\Lambda\simeq 0.003
M_{\rm Pl}$, respectively.  In {}figure~\ref{fig3} we have $Q_0\simeq
1.3\times 10^{-2}$, with $f\simeq 0.035 M_{\rm Pl}$ and $\Lambda\simeq
0.0016 M_{\rm Pl}$.

The results displayed in {}figures~\ref{fig2} and \ref{fig3} show that
the most important effect of including the thermal corrections to the
gauge field background is on the evolution of $\psi$ itself. In the
absence of a thermal mass correction, the gauge field background is
sustained throughout the inflationary evolution.  However, in the
presence of the thermal correction it is eventually driven to zero
well before inflation ends. {}Furthermore, the larger is the initial
value for the dissipation ratio $Q$, the sooner $\psi$ is driven to
zero. {}For the parameters considered, this suppression of $\psi$
already happens in the weak regime of warm inflation, $Q\ll 1$. 
{}For the parameters considered and for a dissipation ratio
$Q \gtrsim 5\times 10^{-2}$, we find that the gauge field background already 
vanishes at the onset and remains null throughout the inflationary
evolution.

In principle, we could believe that the results could be changed by
making $g$ very small, thus suppressing the thermal effects and allowing
a nonvanishing value for $\psi$ to be sustained throughout inflation.
However, this also affects the dissipation coefficient through its dependence
on the coupling $g$, forcing either $\lambda$ to be larger or $f$ to be
smaller. In general we find that larger values of $Q$ becomes hard to be
obtained (we also recall the results of 
refs.~\cite{Montefalcone:2022jfw,Zell:2024vfn} of the difficulties of having
warm inflation in the strong regime, $Q >1$, in axion-like inflaton potentials).
In special we find that the larger is the dissipation ratio $Q$, one
requires smaller couplings to support a gauge background $\psi\neq 0$.

Another approach to favor chromoinflation in the current scenario would be to 
increase the gauge background field. This would require a gauge field mass, $m_A$, 
much larger than the temperature. This would suppress thermal effects from 
eq.~(\ref{VeffT}), as they would be Boltzmann-suppressed.
However, this introduces a new challenge. A larger gauge field mass, resulting 
from a non-zero $\psi$, would also strongly suppress the dissipation coefficient 
in eq.~(\ref{Upsilon}). This is analogous to what is expected to happen
in the electroweak phase transition~\cite{Cohen:1993nk}: above the electroweak 
scale, massless gauge bosons allow rapid sphaleron processes; below the scale, 
mass acquisition suppresses these processes.
Similarly, here, a larger gauge field mass would suppress the dissipation 
necessary for warm inflation. While cold chromoinflation could still occur, 
a warm inflation regime would be precluded.

As discussed at the end of the last section, we can interpret the
vanishing of $\psi$ as a true phase transition. This is explicitly
illustrated in {}figure~\ref{fig4}, where we show snapshots of the
effective potential eq.~(\ref{VeffpsiT}) as a function of $\psi$ at
different values of e-folds for the case of $Q_0\simeq 1.3\times
10^{-2}$, which is the case shown in {}figure~\ref{fig3}. {}From the
results of {}figures~\ref{fig2} and \ref{fig3} (e.g. from the panels (b)
in those figures), this behavior of the minimum of the effective
potential can be interpreted in terms of an increasing value of the
temperature at the corresponding values of e-folds.

The corresponding critical temperature for the (first-order) phase
transition seen in {}figure~\ref{fig4}(b) is $T_c \simeq  1.6 \times
10^{-4} M_{\rm Pl}$, while for the case of the parameters shown in
{}figure~\ref{fig2}(d), i.e. for $Q_0=1.2\times 10^{-3}$ corresponds to
$T_c\simeq 4.9 \times 10^{-4} M_{\rm Pl}$. These values agree well with the
simpler estimate given by eq.~(\ref{Tc}). Hence, the smaller is
$Q_0$, the larger becomes $T_c$. 

The minimum of the effective potential, corresponding to the solution
$\Psi_+$ in eq.~(\ref{Psipm}), as a function of the temperature, is
displayed in {}figure~\ref{fig5}. It shows the behavior typical of a
order parameter as a function of the temperature when the phase
transition is first order. The order parameter (which is here represented by
$\psi_+$) jumps discontinuously from a nonnull to null value across
the phase transition point.

\section{Conclusion} 
\label{sec5}

A pseudo-Nambu-Goldstone scalar field coupled to non-Abelian gauge
fields via a Chern-Simons term provides a successful model of warm
inflation, known as minimal warm inflation. This model arises from the
natural dissipation of the axion-like inflaton field due to sphaleron
transitions in a thermal bath, ensuring a warm inflationary regime
where the temperature $T$ exceeds the Hubble parameter $H$.

Chromoinflation, another model involving axion-like fields coupled to
non-Abelian gauge fields, allows for a homogeneous background gauge
field. A natural extension is to incorporate this into the minimal
warm inflation framework. The thermalized bath of gauge field
fluctuations is expected to induce thermal corrections, including a
thermal plasma mass, for the background gauge field.

In this paper, we investigated the impact of this thermal mass
correction on the background gauge field $\psi$. Our findings reveal
that the evolution of $\psi$ can be significantly influenced by
thermal effects, particularly at larger dissipation ratios $Q$
characteristic of warm inflation. A non-vanishing initial value of
$\psi$ can be rapidly driven to zero by these thermal
effects. Subsequently, the dynamics effectively reduces to that of
minimal warm inflation without a background gauge field.
Conversely, a non-zero gauge field mass, induced by a non-vanishing $\psi$, 
suppresses sphaleron processes responsible for the dissipation term 
in eq.~(\ref{Upsilon}). This suppression hinders warm inflation. 
Therefore, a successful warm inflation scenario in chromoinflation would 
be prevented.

We have demonstrated that the process by which the background gauge
field transitions from a non-zero value to zero closely resembles a
phase transition triggered by temperature variations. At a specific
critical temperature, the gauge field undergoes a first-order phase
transition. We have analytically characterized the properties of this
phase transition. We have also illustrated it by an explicit numerical
example.

In this work, we have primarily focused on the background dynamics. It
is crucial to extend this analysis to perturbations to understand how
they differ from standard warm inflation scenarios without a
background gauge field. The interplay between the background gauge
field and thermal effects could lead to significant
deviations. {}Furthermore, the backreaction effects, commonly studied
in cold
chromoinflation~\cite{Maleknejad:2016qjz,Maleknejad:2018nxz,Ishiwata:2021yne},
should be considered. These additional factors are essential for
comparing the model's predictions with observational data. We plan to
address these aspects in a future work.

\acknowledgments

The authors would like to thank A. Berera and A. Maleknejad
for discussions in the early stages of developing this work.
V.K. would like to acknowledge the McGill University Physics Department 
and Trottier Space Institute for hospitality and partial financial support.
R.O.R. acknowledges financial support by research grants from Conselho
Nacional de Desenvolvimento Cient\'{\i}fico e Tecnol\'ogico (CNPq),
Grant No. 307286/2021-5, and from Funda\c{c}\~ao Carlos Chagas Filho
de Amparo \`a Pesquisa do Estado do Rio de Janeiro (FAPERJ), Grant
No. E-26/201.150/2021.


\end{document}